\documentstyle[bo99,psfig]{article}
\def\fxu{{$\rm ergs\ cm^{-2}\ s^{-1}$}}

\newcommand{\lta}{\lesssim}
\newcommand{\bl}{{BL Lacs\ }}

\def\lesssim{\, 
\lower2truept\hbox{${<\atop\hbox{\raise4truept\hbox{$\sim$}}}$}\,}
\def\gtrsim{\, 
\lower2truept\hbox{${> \atop\hbox{\raise4truept\hbox{$\sim$}}}$}\,}

\title{ASCA/ROSAT Observations of PKS 2316-423: Spectral Properties of a Low
Luminosity Intermediate-type BL Lac Object}
\author{S.-J. Xue$^{1}$ and Y.-H. Zhang$^{2}$} 
\affil{1)Beijing Astronomical Observatory and Beijing Astrophysics Center
  of National Astronomical Observatories, Chinese Academy of Sciences.
E-mail: xue@bac.pku.edu.cn \\ \\
2)International School for Advanced Studies, SISSA/ISAS, 
via Beirut 2-4, 1-34014 Trieste, Italy; E-mail: yhzhang@sissa.it }

\begin{document}

\maketitle

\begin{abstract}
We present the analysis of archival data from ROSAT and ASCA of 
a serendipitous source PKS 2316-423. According to its featureless
non-thermal radio/optical continuum, the object has been assumed
as a BL Lac candidate in the literature. 
PKS 2316-423 was evident variable over the multiple X-ray observations.
Specially, a variable high-energy tail
of the synchrotron radiation is revealed.
X-ray spectral analysis provided 
further evidence of its synchrotron-nature broad-band spectrum with steep and
down-curved shape in the range of 0.1--10 keV, this is general signature 
of a HBL. 
The spectral energy distribution (SED) through radio-to-X-ray yields 
the synchrotron radiation peak at frequency 
{\footnotesize $\nu_p=7.3\times10^{15}$ Hz}, 
with integrated luminosity of
 {\footnotesize $L_{\rm syn}=2.1\times 10^{44}\rm\ ergs\ s^{-1}$}, 
this suggest that PKS 2316-423 is a low luminosity BL Lac object with 
high synchrotron peak frequency. 
Further SED analysis suggest that PKS 2316-423 is a very low 
luminosity ``intermediate'' or high energy peaked BL Lac object. 
Given the unusual low luminosity, the further studies of PKS 2316-423 
might give clues on the evolution properties of \bl\ .

\keywords {BL Lac objects: individual (PKS 2316-423)
-- X-rays: galaxies}
\end{abstract}

\section{Introduction}

Earlier studies of BL Lac objects have shown that the systematic 
differences between radio and X-ray selected BL Lac objects (RBLs vs 
XBLs) can be just attributed to orientation differences. Moreover, BL 
Lac objects have been reclassified by a more accurate way ``low energy'' 
and ``high energy'' peaked BL Lac objects (LBLs vs HBLs) based on the peak 
frequency of synchrotron radiation (e.g. Giommi and Padovani 1994). In 
general, RBLs and XBLs tend to be LBLs and HBLs, respectively. They 
generally represent two distinct extremes of BL Lacs. 
However, recent studies from deeper and larger X-ray survey have shown 
that BL Lac objects tend to exhibit more homogeneous distributions of the 
properties 
(Perlman et al. 1998; Caccianiga et al. 1999; 
Laurent-Muehleisen et al. 1999) rather than previously disparate ones. 
This has resulted in important roles of intermediate BL Lac objects (IBLs) 
in revealing BL Lac mysteries. 

In this paper, we present the X-ray spectral analysis (ROSAT and ASCA  
archival data) and spectral energy distribution (SED) 
of PKS 2316-423, aiming at showing its 
intermediate-BL Lac properties. It 
is a southern radio source at $z=0.0549$, and was formerly classified
as a BL Lac candidate on the base of its featureless
non-thermal radio/optical continuum (Crawford \& Fabian 1994; Padovani \&
Giommi 1996). We noticed this object as it has been the brightest contaminating
source to the nearby narrow-line X-ray galaxy--NGC 7582 (Xue et al. 1998)
 in most of its historical X-ray records. 
 
The ROSAT(PSPC) and ASCA satellites observed this object as a serendipitous 
source in April 1993 and November 1994 respectively. These observations,
together with the two ROSAT/HRI observations made in 1992 and 1993,  
could not only extend our knowledge of the source SED properties to the 
X-ray domain ($\lta10$ keV), but also provide a good opportunity for X-ray 
spectroscopic studies in the range of 0.1--10 keV. Which turn out to 
be very important for the unambiguous classification of the source.

\section{Spectral analysis}

No evident variations in the source
count rate were detected over both observations
spanning about half a day and a little more than one day. Thus
time-averaged spectra from both satellites
were used for Spectral analyzing.

A simple power-law model, with photon index of $\Gamma\sim2.0$ 
and absorption column density at the Galactic value,
gives acceptable fit to the ROSAT/PSPC data (Figure 1).
The inferred intrinsic luminosity is $5.7\times10^{43}\rm\ ergs\ s^{-1}$,
which is similar to that
of other non-quasar AGN. The source was
observed twice, and showed consistent fluxes,
with ROSAT HRI in June 1992 and May 1993 respectively.
However, the brightness decreased by 30\% 
from the later HRI observation to the
PSPC observation which was taken one week apart. These
factors suggest the source is variable and thus there might be non-thermal
origin for the X-ray flux.

A simple power-law model fails to well describe the ASCA data, mainly due to
an abnormal excess absorption above the Galactic value is required.
Consider that this excess absorption 
might be an artifact due to a false spectral model, 
we next fitted the data with a broken power-law with
free break energy. Thus the fit to the data is notablely improved 
at a $\sim90$\% level,  
yields the absorption in consistent with the Galactic value
and two powerlaw components with a break-point at 
$\sim 2.1$ keV (see Table 1). The lower-energy part is flatter 
with a slope in good agreement with that of ROSAT spectrum; the higher-energy 
part is steeper with $\Gamma=2.6^{+0.3}_{-0.3}$.

Comparison with the ROSAT/PSPC observation, the ASCA data indicate 
that the source brightness decreased by 33\% in the 0.1--2.4 keV band
in a 1.5 years interval.  
Meanwhile the broad-band X-ray spectrum
remained the shape at the 
                                                                           
\newpage
\begin{figure}[t]
\psfig{figure=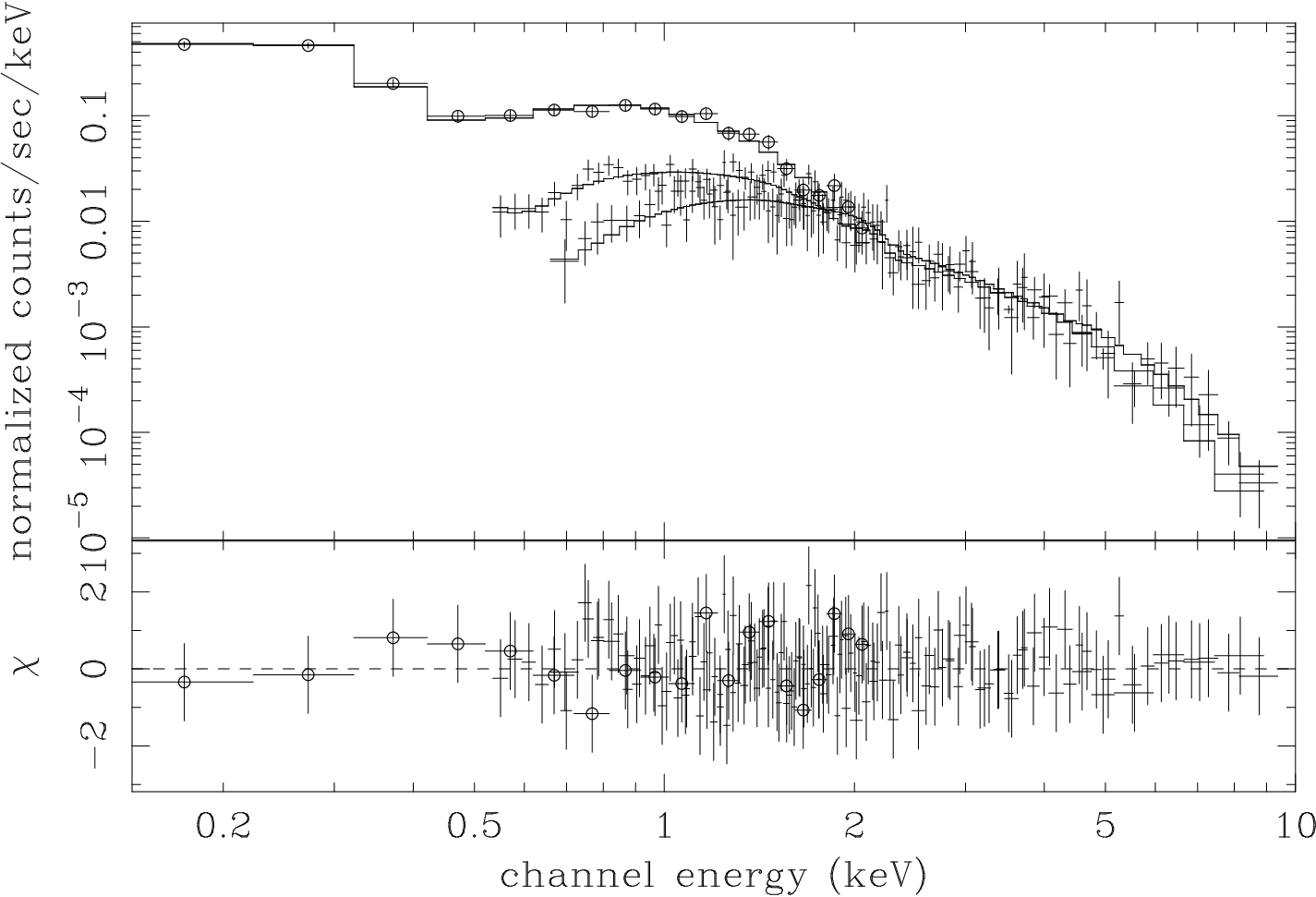,height=4.3cm,angle=270}
\footnotesize
Figure 1.
Folded ROSAT/PSPC and ASCA SIS/GIS spectra of PKS 2316-423.
\end{figure}

\vspace*{-6.2cm}
\begin{flushright}
\begin{minipage}[t]{6.3cm}
\footnotesize
\centerline{Table1. Spectral fitting (with 90\% errors).}
\vspace*{-7pt}

\begin{center}
\begin{tabular}{lccl}
\hline\hline
& & & \\[-8pt]
{Data} & {$N_{\rm H}$} & {$\Gamma_1$}
& {$\chi^{2}_{\nu}$/d.o.f.} \\
& [$10^{20}~{\rm cm^{-2}}$] &  & \\
\hline
& & & \\[-7pt]
ROSAT &  $1.4^{+0.5}_{-0.4}$
& $2.0^{+0.2}_{-0.2}$ & 1.1/17  \\
[2pt]
ASCA &  $2.2^{+2.2}_{-2.0}$
& $2.0^{+0.4}_{-0.2}$ & 1.0/131\\
[2pt]
\hline
\end{tabular}
\\[1pt]
\begin{tabular}{ccc}
\hline
\\[-8pt]
{$\Gamma_2$} & {E$_{\rm break}$}
& {F$_{0.1-2\rm keV}^{obs}$} \\
& keV & ($10^{-12}$\fxu) \\
\hline
\\[-5pt]
-- & --
& $2.63^{+0.15}_{-0.18}$ \\
[2pt]
$2.6^{+0.3}_{-0.3}$ & 2.1 &
$1.35^{+0.33}_{-0.20}$ \\
[3pt]
\hline
\end{tabular}
\end{center}
\end{minipage}
\end{flushright}
\vspace{0.6cm} 

\noindent
lower-energy part, and hardened the slope
in higher-energy range, which was likely in a manner of the prediction of the 
synchrotron radiation losses.

\section{Spectral Energy Distribution}

The composite SED (Figure 2), from both space and ground-based observations,
provides further insights into the object.
It is clear that the  
SED from radio to X-ray is possibly from only one radiation component 
(Synchrotron emission) and peaks at a higher  
frequency falling in the EUV/soft-X-ray band ($\nu_p=7.3\times10^{15}$ Hz).
The optical and ultraviolet 
radiation appear to be a continuation of the radio synchrotron spectrum;
the X-ray data are likely from a common emission origin as the lower 
energy parts and represents a high energy tail of the synchrotron spectrum. 
Other relevant slope parameters from the SED 
are listed in Table 2.

For a comparison, we plotted in Figure 2 
the EGRET sensitivity threshold as an upper limit to the GeV flux 
(marked by an arrow), since the source was never detected at $\gamma$-ray. 
It is shown that the source is well dominated by a synchrotron process.

\section{Discussions}

Putting 2316-423 on the $\alpha_{ro}$ vs $\alpha_{ox}$ color-color 
diagram, we  
find it is in the intermediate range of BL Lacs. As we know, 
$\alpha_{XOX}$ can more precisely measure spectral changes from optical to 
soft X-ray bands, however, the values of $\alpha_{XOX}$ for PKS 2316-423 
depend on the assumption of the X-ray spectral indices, being 0.18/0.26 
and -0.42/-0.34 for $\alpha_x=1.0$ and 1.6, respectively. These 
values should locate in the intermediate range of the $\alpha_{XOX}$ 
distribution of recent BL Lacs samples (Laurent-Muehleisen et al. 1999).

The importance of the frequency at which the synchrotron radiation 
peaks is that it provides a powerful diagnostics for the physical 
condition of the emitting region. Recent studies showed that among BL 
Lacs the synchrotron peak frequencies are inversely correlated with their 
luminosities (Fossati et al. 1998). We could put 

\newpage
\begin{figure}[t]
\psfig{figure=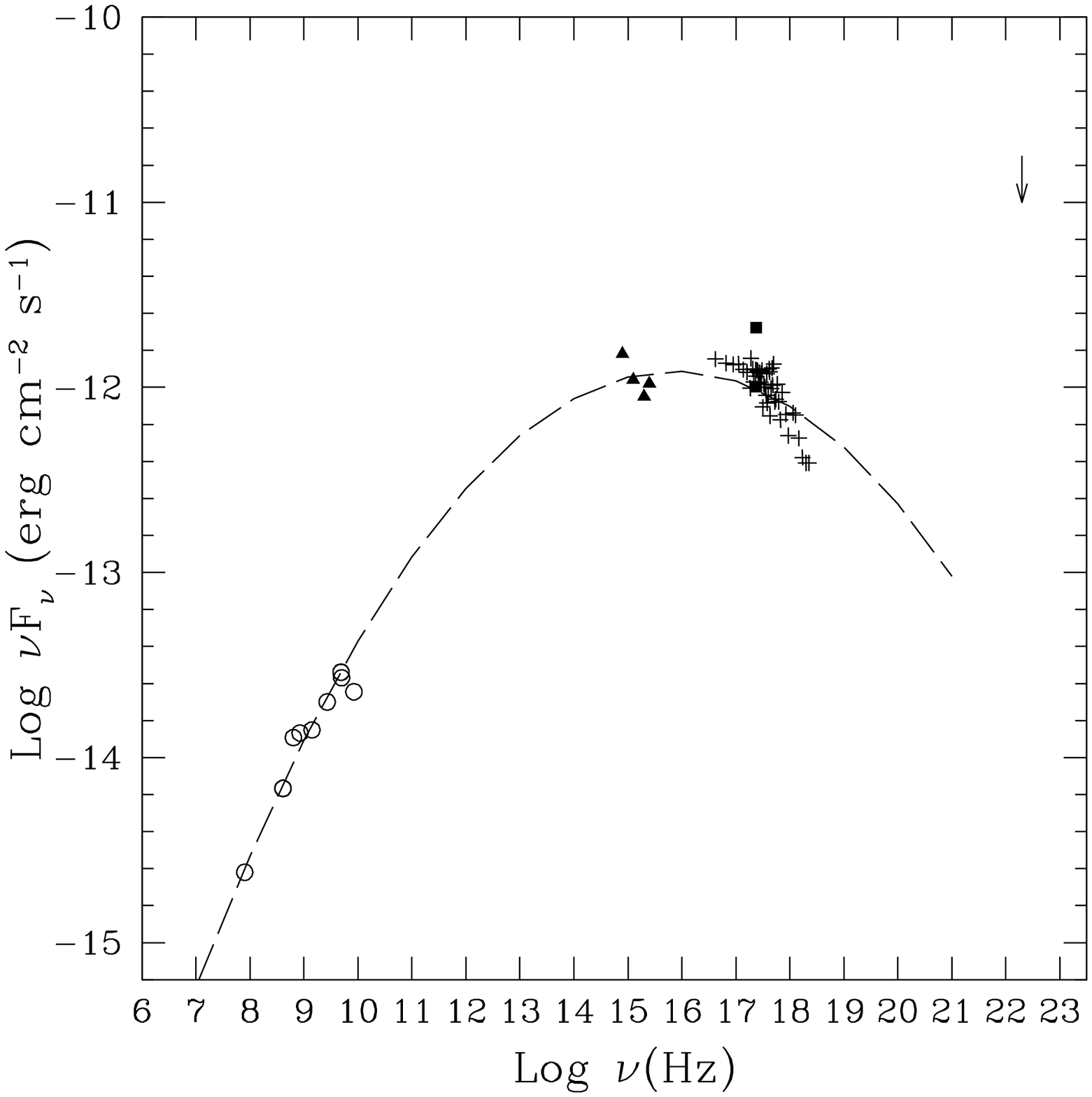,height=5cm}
\footnotesize
Figure 2. The multifrequency SED of PKS 2316-423 and its parabolic fit. The 
X-ray points are data from this paper, plotted with solid squares 
for ROSAT/HRI and cross marks for ASCA and ROSAT/PSPC.
The UV/optical points are data from Crawford and Fabian (1994) plotted 
with solid triangle. Circle symbols represent radio data
from NASA/IPAC Extra-galactic Database (NED).
\end{figure}

\vspace*{-6.5cm}
\begin{flushright}
\begin{minipage}[t]{6.5cm}
\footnotesize
\centerline{Table 2. Broad-band slopes of SED }
\vspace*{7pt}
\begin{center}
\begin{tabular}{cccc}
\hline\hline
& & & \\[-5pt]
{$\alpha_{ro}$} & {$\alpha_{ox}$} & {$\alpha_{x}$}
& {$\alpha_{x}$} \\
& & (0.1-2 keV) & (2-10 keV) \\
\hline
& & & \\[-7pt]
0.56 & 1.18/1.26 & 1.0 & 1.6 \\
[2pt]
\hline
\end{tabular}
\end{center}
\end{minipage}
\end{flushright}
\vspace{3cm}

\noindent
PKS 2316-423 on the 
Figure 7c of Fossati et al. (1998). Due to its lowest peak luminosity, 
PKS 2316-423 should locate the right-bottom end, this means that the peak 
frequency of PKS 2316-423 would be around $\sim 10^{18}$ Hz, however, our fit 
to the SED just gives $\nu_p \sim 10^{16}$ Hz. Therefore, we suggest that 
PKS 2316-423  might be a low luminosity ``intermediate'' object between HBLs 
and LBLs. 

In a summary, the X-ray spectral and SED analysis of PKS 2316-423 point 
out its IBL or HBL attributes with very low luminosity compared with the 
most recent BL Lac samples. Because of its peculiar low 
luminosity, however, the more detailed studies of 
PKS 2316-423 will shed light on the evolution of BL Lac objects.    

\begin{acknowledgements}
This research has 
made use of the NASA/IPAC Extra-galactic Database (NED) which is 
operated by the Jet Propulsion Laboratory, California Institute 
of Technology, under contract with the National 
Aeronautics and Space Administration. S.J.X. acknowledges the 
financial support from Chinese Post Doctoral Program.
\end{acknowledgements}


\begin{references}

\ref Caccianiga, A., Maccacaro, T., Wolter, A. et al., 1999, ApJ, 513, 51
\ref Crawford C.S., Fabian A.C., 1994, MNRAS, 266, 669
\ref Fossati, G., Maraschi, L., Celloti, A. et al., 1998, MNRAS, 299, 433
\ref Giommi P., Padovani P., 1994, MNRAS, 268, L51
\ref Laurent-Muehleisen  et al., 1999, astro-ph/9905133
\ref Perlman, E.S., Padovani, P., Giommi, P., 1998, AJ, 115, 1253
\ref Padovani P., Giommi P., 1995, ApJ, 444, 567
\ref Padovani P., Giommi P., 1996, MNRAS, 279, 526
\ref Xue S.J., Otani C., Mihara T. et al. 1998, PASJ, 50, 519

\end{references}
\end{document}